\newcounter{myctr}
\def\myitem{\refstepcounter{myctr}\bibfont\noindent\ifnum\themyctr>9\else\phantom{0}\fi\hangindent17pt\themyctr.\enskip}

\documentclass{ws-ijqi}

\begin{document}

\markboth{M. Sun {\it et al.}}
{Security of New Two-Way Continuous-Variable Quantum Key Distribution Protocol}

%%%%%%%%%%%%%%%%%%%%% Publisher's Area please ignore %%%%%%%%%%%%%%
\catchline{}{}{}{}{}
%%%%%%%%%%%%%%%%%%%%%%%%%%%%%%%%%%%%%%%%%%%%%%%%%%%%%%%%%%%%%%%%%%%

\title{SECURITY OF A NEW TWO-WAY CONTINUOUS-VARIABLE QUANTUM KEY DISTRIBUTION PROTOCOL}

\author{MAOZHU SUN, XIANG PENG$^*$, YUJIE SHEN and HONG GUO$^\dag$}

\address{CREAM Group, The State Key Laboratory of  Advanced Optical Communication Systems and Networks and Institute of Quantum Electronics, School of Electronics Engineering and Computer Science, Peking University, Beijing 100871, China\\
$^*\!$xiangpeng@pku.edu.cn\\$^\dag$hongguo@pku.edu.cn}

\maketitle

\begin{history}
\received{Day Month Year}
\revised{Day Month Year}
%\accepted{Day Month Year}
%\comby{(xxxxxxxxxx)}
\end{history}

\begin{abstract}
The original two-way continuous-variable quantum-key-distribution (CV QKD) protocols [S. Pirandola, S. Mancini, S. Lloyd, and S. L. Braunstein, Nature Physics \textbf{4}, 726 (2008)] give the security against the collective attack on the condition of the tomography of the quantum channels. We propose a family of new two-way CV QKD protocols and prove their security against collective entangling cloner attacks without the tomography of the quantum channels. The simulation result indicates that the new protocols maintain the same advantage as the original two-way protocols whose tolerable excess noise surpasses that of the one-way CV-QKD protocol. We also show that all sub-protocols within the family have higher secret key rate and much longer transmission distance than the one-way CV-QKD protocol for the noisy channel.
\end{abstract}


\section{Introduction}	%) A SECTION HEADING

Quantum key distribution is well applied to cryptography due to its unconditional security based on quantum mechanics.\cite{Scarani} In particular, continuous-variable quantum key distribution (CV QKD) has attracted much attention in recent years because it has potentially faster and more efficient detection and key rate than single-photon QKD.\cite{1Grosshans05,8Pirandola,P. K. Lam no switching,het} One-way CV QKD allows the quantum state to pass through the channel only from the sender (Alice) to the receiver (Bob), which brings a limitation that the channel loss is no more than 3 dB in direct reconciliation.\cite{2Grosshans02} Although the post-selection\cite{postselection Silberhorn} or the reverse reconciliation\cite{Grosshans03 rev entangle cloner,11Grosshans03} overcomes this drawback, the secret key rate is strongly affected by excess noise.\cite{6Pirandola} To enhance the tolerable excess noise, the two-way CV-QKD protocols are proposed to go beyond the 3 dB limit and meanwhile tolerate more excess noise than one-way protocols.\cite{6Pirandola,8Pirandola}

The procedure of implementing the original two-way CV protocol is briefly introduced below. The entanglement-based (EB) scheme of a sub-protocol in the original two-way protocols, $\textrm{Het}^\textit{2}$ protocol, is shown in Fig. 1(a), and can be described as:\cite{6Pirandola,8Pirandola}

\emph{Step one}. Bob initially prepares an EPR pair with variance $V$ and keeps one mode $B_1$ while sending the other mode $C_1$ to Alice through the channel where Eve may perform her attack.

\emph{Step two}. Alice encodes her information by applying a random phase-space displacement operator $D(\alpha)$ to her received mode $A_{in}$ and then sends the mode $A_{out}$ back to Bob through the channel. Note that $\alpha=(Q_A+iP_A)/2$, and $Q_A$ or $P_A$ has a random Gaussian modulation with the variance of $V-1$, respectively.

\emph{Step three}. Bob measures both his original mode $B_1$ and received mode $B_2$ with heterodyne detection to get the variables $x_{B_{1X}}$ and $p_{B_{1P}}$ as well as $x_{B_{2X}}$ and $p_{B_{2P}}$, respectively.

\emph{Step four}. Alice and Bob implement the postprocessing including reconciliation and privacy amplification.\cite{postprocessing} In this procedure, Bob needs to combine both outcomes from $B_1$ and $B_2$ to construct the optimal estimator to Alice's corresponding variables \{$Q_A$, $P_A$\}. After the steps above, Alice and Bob can share a string of identical key that Eve does not know.

However, to analyze the security under general collective attack, the original two-way protocols need to construct the hybrid protocol where Alice randomly switches between one-way (switch OFF, where Alice detects the incoming mode and sends a new state back to Bob) and two-way schemes (switch ON) for implementing the tomography of the quantum channels and for both parameter estimation and key distribution,\cite{6Pirandola,8Pirandola} as shown in Fig.1. This hybrid scheme increases the complexity in a real setup. Moreover, it is difficult to implement the tomography of quantum channels in a real experiment. In this paper, we modify the original two-way protocol by replacing the displacement operation and the ON-OFF switch with a passive operation on Alice's side, and give a feasible prepare-and-measure (PM) scheme, which pushes the two-way protocol to be easily applied in practice. Considering that Gaussian collective attack is optimal, we will prove the security of the new protocol under collective entangling cloner attacks which are a special case of general Gaussian collective attack thoroughly researched in Ref.~\refcite{channel S. Pirandola,channel A. S. Holevo}. This paper is organized as follows. Section 2 contains the statements of our new two-way CV-QKD protocol. In Sec. 3, we give a theoretical  analysis of the security of the new two-way CV-QKD protocol against Gaussian collective attack by using the optimality of Gaussian collective attack. In Sec. 4, we investigate the numerical simulation of the secret key rate under collective entangling cloner attacks. Finally, in Sec. 5, we conclude our results and indicate some open questions.

\section{A New Two-Way CV QKD Protocol}

We modify the original two-way protocols by replacing the displacement operation and the ON-OFF switch with the passive operation on Alice's side. The EB scheme of $\textrm{Het}^\textit{2}_{\textrm{M}}$ protocol after modifying the $\textrm{Het}^\textit{2}$ protocol is shown in Fig. 1(b). In $\textrm{Het}^\textit{2}_{\textrm{M}}$, the second and fourth steps of $\textrm{Het}^\textit{2}$ are changed into

\emph{Step $two^\prime$}. With using a beam splitter (transmittance: $T_A$), Alice couples one mode of another EPR pair (variance: $V_A$) with the received mode $A_{in}$ from Bob and sends the coupling mode $A_{out}$ back to Bob. She also measures the other mode $A_1$ of this EPR pair with heterodyne detection to get the variables \{$x_{A_{1X}}$, $p_{A_{1P}}$\} and randomly measures the position quadrature $x$ or the momentum quadrature $p$ of the coupling mode $A_2$ from the beam splitter with homodyne detection.

\emph{Step $four^\prime$}. Alice and Bob implement the postprocessing including the reconciliation and privacy amplification.\cite{postprocessing} In this procedure, the homodyne detection on the mode $A_2$ is used to estimate the channel's parameters and Bob uses $x_{B}=x_{B_{2X}}-kx_{B_{1X}}$ and $p_{B}=p_{B_{2P}}+kp_{B_{1P}}$ to construct the optimal estimator to Alice's corresponding variables \{$x_{A_{1X}}$, $p_{A_{1P}}$\}, where $k$ is the channel's total transmittance which is obtained by reconciliation. The other steps of $\textrm{Het}^\textit{2}_{\textrm{M}}$ are the same as those of $\textrm{Het}^\textit{2}$.

In Fig. 1(b), Alice's beam splitter $T_A$ couples the two uncorrelated states respectively from Alice and Bob. The action of the beam splitter $T_A$ is equivalent to a unitary transformation. One output mode $A_2$ of this beam splitter is kept and measured on Alice's side and the other mode $A_{out}$ is sent to Bob though the channel. The effects of system parameters and environment parameters on entanglement are discussed in detail in Ref.~\refcite{G. Zen}. Here the two channels affect the entanglement degrees of those three pairs of states: $B_1$ and $A_2$, $B_1$ and $B_2$, and $A_1$ and $B_2$. The effect on the channels can be ascribed to the action of Eve. Considering one-mode Gaussian attack, the two channels can be described as two independent Gaussian-Entangling-Cloner attacks.\cite{6Pirandola} It is necessary to estimate the channel's parameters by the measurement values of Alice and Bob in security analysis.

The PM scheme of $\textrm{Het}^\textit{2}_{\textrm{M}}$ protocol is shown in Fig. 1(c), which is equivalent to the EB scheme in Fig. 1(b).\cite{Grosshans03 rev entangle cloner} In Fig 1(c), with using the random numbers $m$ and $n$, Bob randomly modulates the amplitude ($\textrm{A}$) and the phase ($\rm{\phi}$) of the coherent state from his laser source (LS1), and then sends the state to Alice. Alice's laser source (LS2) is coherent with Bob's LS1 by phaselock and time synchronization techniques.\cite{phase lock} Similar to Bob's modulation, Alice uses two other random numbers $r$ and $s$ to encode information. After that, the beam splitter (transmittance: $T_A$) couples Alice's signal with the signal from Bob's side, and outputs one mode back to Bob and another mode measured with homodyne detection. At last, the returned mode is measured with heterodyne detection on Bob's side. Note that the local oscillator and the switch which randomly controls the homodyne detection to detect the $x$ or $p$ quadrature are omitted for concision in Fig. 1.

In addition, the other original\cite{6Pirandola} (e.g., $\textrm{Hom}^\textit{2}$) can be modified to new protocol (e.g., $\textrm{Hom}^\textit{2}_{\textrm{M}}$) by changing the displacement into the coupling of the EPR pair, correspondingly. According to Bob's detection, we also propose a new sub-protocol $\textrm{Hom}$-$\textrm{Het}_\textrm{M}$ ($\textrm{Het}$-$\textrm{Hom}_\textrm{M}$ ) where Bob measures his mode $B_1$ with homodyne (heterodyne) detection and measures his mode $B_2$ with heterodyne (homodyne) detection.

\begin{figure}[pb]
\centerline{\psfig{file=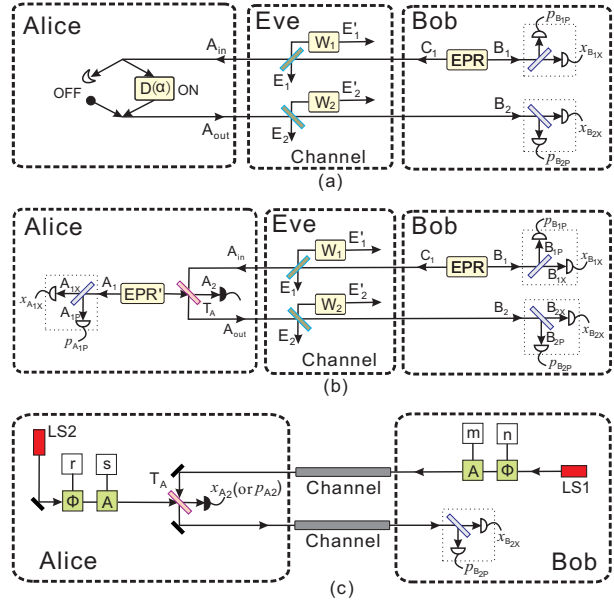,width=8cm}}
\vspace*{8pt}
\caption{(a) The EB scheme of hybrid $\textrm{Het}^\textit{2}$ protocol. Bob measures one half of the EPR pair (EPR) with heterodyne detection and sends the other half to Alice. After through the path switch ON or OFF on Alice's side, the back state $B_2$ is measured with heterodyne detection. There are two independent Gaussian-Entangling-Cloner attacks (with variances $W_1$ and $W_2$) on the channels whose transmittances are modeled by two beam splitters. The letters (e.g., $\textrm{B}_1$) beside arrows: the mode at the corresponding position; crescent: detection; the circle: new state; the dashed box at $B_1$ and $B_2$: the heterodyne detection.
(b) The EB scheme of $\textrm{Het}^\textit{2}_{\textrm{M}}$ protocol. It is the same as (a) on Bob's side. On Alice's side, Alice measures one mode of her EPR pair ($\rm{EPR}^\prime$) with heterodyne detection and measures  one mode from a beam splitter with the transmittance $T_A$ by homodyne detection. The other mode from this beam splitter is returned back to Bob. (c) The PM scheme of $\textrm{Het}^\textit{2}_{\textrm{M}}$ protocol. Bob sends a coherent state to Alice, then measures the back state with heterodyne detection to get the position  ($x_{B_{2X}}$) and the momentum  ($p_{B_{2P}}$) quadratures. Alice gets another value $x_{A_2}$ by the homodyne detection. LS1 and LS2: laser source; A: amplitude modulator; $\phi$: phase modulator; m, n, r and s: random number generator.}
\end{figure}

\begin{figure}[pb]
\centerline{\psfig{file=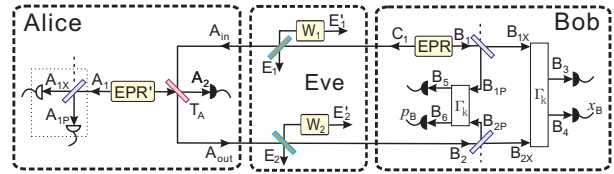,width=8cm}}
\vspace*{8pt}
\caption{The equivalent scheme to Fig. 1 (b). Bob uses two unitary transformations $\Gamma_k$ to change the modes $B_{2X}$ and $B_{1X}$ ($B_{2P}$ and $B_{1P}$) into $B_3$ and $B_4$ ($B_5$ and $B_6$), where $\Gamma_k$ is a CV C-NOT gate. By homodyne detection on the position (momentum) quadrature of $B_4$ ($B_6$), $x_B$ ($p_B$) is obtained. The dashed line into beam splitter: vacuum state.}
\end{figure}

\section{A Theoretical Analysis of the Security of the New Two-Way Protocol}

We consider the EB scheme of $\textrm{Het}^\textit{2}_{\textrm{M}}$ protocol in reverse reconciliation. The secret key rate is\cite{KRB1 J. Lodewyck,KRB2A. Leverrier}
\begin{equation}\label{KR}
K_R=\beta I_{BA}-I_{BE},
\end{equation}
where $\beta$ is the reconciliation efficiency, $I_{BA}$ is the mutual information between Alice and Bob, $I_{BE}$ is the mutual information between Eve and Bob.

According to \emph{step $four^\prime$}, in Fig. 1(b), $I_{BA}=\log_2\left(V_{A^M}/V_{A^M|B}\right)$, where $V_{A^M}$ and $V_{A^M|B}$ are Alice's variance and conditional variance on Bob, respectively.\cite{11Grosshans03} $I_{BA}$ can be obtained with Alice's and Bob's data. As far as $I_{BE}$ is concerned, according to Holevo bound,\cite{13Holevo} we get
\begin{eqnarray}\label{IBE}
    &&I_{BE}=S(E)-S(E|x_B,p_B),
\end{eqnarray}
where $S(E)$ is Eve's von Neumann entropy and $S(E|x_B,p_B)$ is Eve's conditional von Neumann entropy on Bob's data.

Because the calculation of $S(E|x_B,p_B)$ relates to Bob's postprocessing, in order to obtain the secret key rate, Fig. 2 instead of Fig. 1(b) is used for security analysis. In Fig. 2, Bob applies two unitary transformations $\Gamma_k$ to the modes $B_{2X}$ and $B_{1X}$ as well as to the modes $B_{1P}$ and $B_{2P}$ , respectively, in order to get $x_B$ ($p_B$) by measuring the position (momentum) quadrature of $B_4$ (or $B_6$). Note the order of the transformation, e.g., $(x_{B_4},p_{B_4},x_{B_3},p_{B_3})^T=\Gamma_k (x_{B_{2X}},p_{B_{2X}},x_{B_{1X}},p_{B_{1X}})^T$, where $x_{B_4}$, $p_{B_4}$, $x_{B_3}$ and $p_{B_3}$ are the $x$ and $p$ quadratures of the modes $B_3$ and $B_4$ and $\Gamma_k$ is a continuous-variable C-NOT gate\cite{QND,doctor Raul,Quantum Computation and Quantum Information Nielsen}
\begin{equation}
 \Gamma_k=\left(
\begin{array}{cccc}
1&0&-k&0\\
0&1&0&0\\
0&0&1&0\\
0&k&0&1
\end{array}
\right).
\end{equation}
Considering the assumption that Eve has no access to the interior of Bob,\cite{Scarani} Eve obtains the information only  from Bob's input and output. Because the unitary transformation $\Gamma_k$ doesn't change the von Neumann entropy of the system\cite{doctor Raul} $B_{2X}B_{1X}B_{2P}B_{1P}A_2A_{1X}A_{1P}E$ and the variables $x_B$ and $p_B$ are same to both Figs. 1(b) and 2, Eve's von Neumann entropy and conditional von Neumann entropy on Bob in Fig. 2 are equivalent to those in Fig. 1(b). A detailed proof can be seen in Appendix A. In addition, taking into account that $I_{BA}$ is the same for both systems, the secret key rate is same to both Figs. 1(b) and 2. Thus, we use Fig. 2 to analyze the security in the following.

First, we show that the Gaussian attack is optimal to the new protocol. According to the \emph{step $two^\prime$} and \emph{step three} of the protocol $\textrm{Het}^\textit{2}_{\textrm{M}}$, Alice and Bob measure the mode $A_2$ with homodyne detection and measure the modes $A_1$, $B_1$ and $B_2$ with heterodyne detection. This is equivalent to the scheme that Alice and Bob measure the mode $A_1$ with heterodyne detection and measure the modes $A_2$, $B_3$, $B_4$, $B_5$ and $B_6$ with homodyne detection in Fig. 2, i.e., Alice and Bob measure all modes except Eve's modes. In Fig. 2, $\rho_E$, $\rho_B$ and $\rho_A$ denote the states of Eve, the modes $B_4B_6$ and the modes $A_2A_{1X}A_{1P}B_3B_5$, respectively. It is easily seen that $\psi_{ABE}$ is a pure state and $\rho_{AB}$ is the purification of $\rho_E$. Because Alice and Bob's heterodyne or homodyne detection on their modes does not mix the $x$ and $p$ quadratures and Alice and Bob use the second-order moments of the quadratures to calculate the secret key rate bound, the new protocol can satisfy the requirement of the optimality of Gaussian collective attack.\cite{doctor Raul} Thus, when the corresponding covariance matrix $\Gamma_{AB}$ of $\rho_{AB}$ is known and fixed for Alice and Bob, the Gaussian attack is optimal.\cite{14N. J. Cerf2006,optimal M. Navascues,optimal M. M. Wolf,optimal Leverrier} Therefore, Eve's accessible information can be bounded by only considering Eve's Gaussian collective attack. In the following part, $I_{BE}$ is calculated using some ideas proposed in Ref.~\refcite{Q-Y. Cai}.

Second, to calculate $S(E)$, one needs to know $S(\rho_{AB})$ because  $\psi_{ABE}$ is a pure state and $S(E)=S(\rho_{AB})$. The entropy $S(\rho_{AB})$ of the Gaussian state $\rho_{AB}$ is calculated according to its corresponding covariance matrix $\Gamma_{AB}$. Note that
\begin{equation} \label{rho_AB}
  \Gamma_{AB}=\left[\Gamma_k\oplus\Gamma_k\oplus\mathbb{I}_3\right]\Gamma_{B_{2X}B_{1X}B_{1P}B_{2P}A_2A_{1X}A_{1P}}\left[\Gamma_k\oplus\Gamma_k\oplus\mathbb{I}_3\right]^T,
\end{equation}
where $\mathbb{I}_3$ is a $6\times6$ identity matrix and $\Gamma_{B_{2X}B_{1X}B_{1P}B_{2P}A_2A_{1X}A_{1P}}$ is the corresponding covariance matrix of the state $B_{2X}B_{2P}B_{1X}B_{1P}A_2A_{1X}A_{1P}$ or (seen in Appendix B)
\begin{eqnarray}\label{Gamma_C}
\Gamma\!\!_{B_{2\!X}\!B_{2\!P}B_{1\!X}\!B_{1\!P}A_2A_{1\!X}\!A_{1\!P}}\!\!=\!\!\!
\left(\!
\begin{array}{ccccccc}
 \gamma_{B_{2X}} & \mathbb{I}\!-\!\gamma_{B_{2X}} & C_1 & -C_1&  C_2 & C_3 & -C_3 \\
 \mathbb{I}\!-\!\gamma_{B_{2X}} &  \gamma_{B_{2P}}  & -C_1 & C_1 & -C_2  & -C_3 & C_3 \\
 C_1 & -C_1 & \frac{1+V}{2}\mathbb{I}  & \frac{1-V}{2}\mathbb{I}  & C_4 & 0  & 0  \\
 -C_1 & C_1 & \frac{1-V}{2}\mathbb{I}  & \frac{1+V}{2}\mathbb{I}  & -C_4 & 0  & 0  \\
 C_2 & -C_2  & C_4 & -C_4 & \gamma_{A_2}  & C_5 & -C_5\\
 C_3 & -C_3 & 0  &   0 & C_5 & \frac{1+V_A}{2}\mathbb{I}  & \frac{1-V_A}{2}\mathbb{I}  \\
 -C_3 & C_3 & 0  & 0  & -C_5 & \frac{1-V_A}{2}\mathbb{I}  & \frac{1+V_A}{2}\mathbb{I}
\end{array}
\!\right),
\end{eqnarray}
in which $\mathbb{I}$ is a $2\times2$ identity matrix. In Eq. (\ref{Gamma_C}), the diagonal elements correspond to the variances of $x$ and $p$ quadratures of the modes $B_{2X}$, $B_{2P}$, $B_{1X}$, $B_{1P}$, $A_2$, $A_{1X}$ and $A_{1P}$ in turn, e.g., $\gamma_{B_{2X}}=diag(\langle x_{B_{2X}}^2\rangle, \langle p_{B_{2X}}^2\rangle)$, and the nondiagonal elements correspond to the covariances between modes, e.g., $C_2=diag(\langle x_{B_{2X}}x_{A_2}\rangle,\langle p_{B_{2X}}p_{A_2}\rangle)$, where $x_{B_{2X}}$, $p_{B_{2X}}$, $x_{A_2}$ and $p_{A_2}$ are the $x$ and $p$ quadratures of the modes $B_{2X}$ and $A_2$, respectively. In experiment, the covariance matrix Eq. (\ref{Gamma_C}) can be calculated by the reconciliation in which Alice and Bob reveal some randomly chosen measurement values obtained by heterodyne detection on the modes $B_2$, $B_1$, $A_1$ and homodyne detection on the mode $A_2$. Note that the $x$ and $p$ quadratures are simultaneously obtained in the heterodyne detection, but Alice needs to randomly measure the $x$ or $p$ quadrature of the mode $A_2$ to obtain the corresponding values of the $x$ and $p$ quadratures of the mode $A_2$. Therefore, Eve's entropy\cite{Holevo S(i)}
\begin{equation}\label{SE}
S(E)=\sum^7_{i=1} G(\lambda _i)=\sum^7_{i=1} G\left(f_{\lambda _i}(\alpha_{mn})\right),
\end{equation}
where
\begin{equation}
G(\lambda _i)=\frac{\lambda _i+1}{2}\log\frac{\lambda _i+1}{2}-\frac{\lambda _i-1}{2}\log\frac{\lambda _i-1}{2},
\end{equation}
and $\lambda_i=f_{\lambda _i}(\alpha_{mn})$ is the symplectic eigenvalue of $\Gamma_{AB}$ which is the function of the element $\alpha_{mn}$ of $\Gamma_{AB}$, seen in Appendix C.

Third, $S(E|x_B,p_B)=S(B_3B_5A_2A_{1X}A_{1P}|x_B,p_B)$ because the state $B_3B_5A_2A_{1X}A_{1P}E$ is a pure state when Bob gets $x_B$ and $p_B$ by measuring the modes $B_4$ and $B_6$. The corresponding covariance matrix $\Gamma_{B_3B_5A_2A_{1X}A_{1P}}^{x_B,p_B}$ of the state $B_3B_5A_2A_{1X}A_{1P}$ conditioned on $x_B$ and $p_B$ can be obtained from $\Gamma_{AB}$\cite{doctor Raul,J. Eisert conditional matrix}
\begin{eqnarray} \label{Gamma(AB)XB}
  \Gamma_{\!\!B\!_3B\!_5A_2A_{\!1\!X}\!A_{\!1\!P}}^{x_B,p_B}\!\!=\!\!\Gamma_{\!\!B_3\!B_5\!A_2\!A_{\!1\!X}\!A_{\!1\!P}}\!\!-\!C_{\!B_4}[X_x\gamma_{B_4}X_x]^{\!M\!P}C_{\!B_4}\!\!\!\!\!^T\!-\!C_{\!B_6}[X_p\gamma_{B_6}X_p]\!^{M\!P}C_{\!B_6}\!\!\!\!\!^T~\!,
\end{eqnarray}
where $\Gamma_{B_3B_5A_2A_{1X}A_{1P}}$, $\gamma_{B_4}$ and $\gamma_{B_6}$ are the corresponding reduced matrixes of state $B_3B_5A_2A_{1X}A_{1P}$, $B_4$ and $B_6$ in $\Gamma_{AB}$, respectively, $C_{B_4}$ and $C_{B_6}$ are their correlation matrixes, $X_x=diag(1,0)$, $X_p=diag(0,1)$ and $MP$ denotes the inverse on the range. Similar to Eq. (\ref{SE}), we obtain
\begin{equation}\label{S(E|x_B,p_B)}
S(E|x_B,p_B)=\sum^5_{i=1} G(\lambda _i^\prime)=\sum^5_{i=1} G\left(f_{\lambda _i^\prime}(\alpha_{mn}^\prime)\right),
\end{equation}
where $\lambda_i^\prime=f_{\lambda _i^\prime}(\alpha_{mn}^\prime)$ is the symplectic eigenvalue of $\Gamma_{B_3B_5A_2A_{1X}A_{1P}}^{x_B,p_B}$ which is the function of the element $\alpha_{mn}^\prime$ of $\Gamma_{B_3B_5A_2A_{1X}A_{1P}}^{x_B,p_B}$, seen in Appendix C.

By substituting Eqs. (\ref{SE}) and (\ref{S(E|x_B,p_B)}) into Eq. (\ref{KR}), the secret key rate is obtained
\begin{equation} \label{K_R(a_{mn})}
K_R=\beta \log_2\frac{V_{A^M}}{V_{A^M|B}}-\sum^7_{i=1} G\left(f_{\lambda _i}(\alpha_{mn})\right)+\sum^5_{i=1} G\left(f_{\lambda _i^\prime}(\alpha_{mn}^\prime)\right).
\end{equation}
In experiment, Alice and Bob can calculate every element of Eq. (\ref{Gamma_C}) according to the measurement values of the modes $B_2$, $B_1$, $A_1$ and $A_2$, then calculate $\alpha_{mn}$ and $\alpha_{mn}^\prime$ from Eqs. (\ref{rho_AB}) and (\ref{Gamma(AB)XB}). Therefore, according to Eq. (\ref{K_R(a_{mn})}), Eve's accessible information under Gaussian collective attacks is bounded and the secret key rate is obtained. Similarly, the security of other sub-protocols of the new two-way CV QKD can be analyzed.

In theory, for the security analysis, we consider collective entangling cloner attacks.  Collective entangling cloner attacks are a specific case of collective Gaussian attacks where the communication channel is linear with transmittance $T$ ($0<T<1$) and thermal noise.\cite{channel S. Pirandola,channel A. S. Holevo} The assumption of linear channel is often used since the linear channel is common in real experiment and easy to be numerically simulated.

To get the elements of Eq. (\ref{Gamma_C}) for numerical simulation, we assume that the two channels are linear with the transmittances $T_1$ and $T_2$ and the noises referred to the input $\chi_1=\varepsilon_1+(1-T_1)/T_1$ and $\chi_2=\varepsilon_2+(1-T_2)/T_2$, respectively, where  $\varepsilon_1$ and $\varepsilon_2$ are the channel excess noises referred to the input. We can obtain
\begin{align} \label{C1-C5}
&\gamma_{B_{2X}}=\gamma_{B_{2P}}=\frac{1}{2}\left\{1 + T\!_2 (V\!_A - T\!_A V\!_A + T\!_1 T\!_A (V + \chi_1) + \chi_2)\right\}  \mathbb{I},\nonumber \\
&\gamma_{A_2}=\left[T\!_A V\!_A + T\!_1 (1- T\!_A) (V + \chi_1)\right]  \mathbb{I},\nonumber  \\
&C_2=\!\!\sqrt{\frac{1}{2}T\!_2 (1- T\!_A) T\!_A} [V\!_A - T\!_1 (V + \chi_1)]  \mathbb{I},  \nonumber \\
&C_1\!=\!\!\frac{1}{2}\! \sqrt{T\!_1T\!_2T\!_A \left(V^2-1\right)}  \sigma_{\!z}, ~~~~~~~~C_3\!=\!\!\frac{1}{2}\!\sqrt{T\!_2\!\left(1\!\!- \!T_{\!\!A}\right)\left(V_{\!\!A}^2\!-\!1\right)}\sigma_{\!z}, \nonumber \\
&C_4\!=\!-\!\sqrt{\!\frac{1}{2}T\!_1\!\left(1\!-\! T_{\!\!A}\right)\left(V^2\!-\!1\!\right)}\sigma_{\!z},~~~~~ C_5\!=\!\!\sqrt{\frac{1}{2}T\!_A \left(V_A^2-1\right)}\sigma_{\!z},
\end{align}
and
\begin{equation}\label{IBA}
I_{BA}=\log_2 \frac{1 + T_1T_2T_A(1 + F) + T_2(V_A - T_AV_A + \chi_2)}{1 + T_1T_2T_A (1 + F) + T_2(1 - T_A + \chi_2)},
\end{equation}
where
\begin{eqnarray}
F=2V - 2\sqrt{V^2-1} + \chi_1,~~~~~~\sigma_z=\left(\begin{array}{cc}
 1&0\\
 0&-1\\
\end{array}\right).
\end{eqnarray}
Substituting above equations into Eq. (\ref{K_R(a_{mn})}), the secret key rate of $\textrm{Het}^\textit{2}_M$ protocol against collective entangling cloner attacks can be obtained. Similarly, the secret key rate of the other sub-protocols of the new two-way CV QKD can be also obtained (seen in Appendix D).

\begin{figure}[pb]
\centerline{\psfig{file=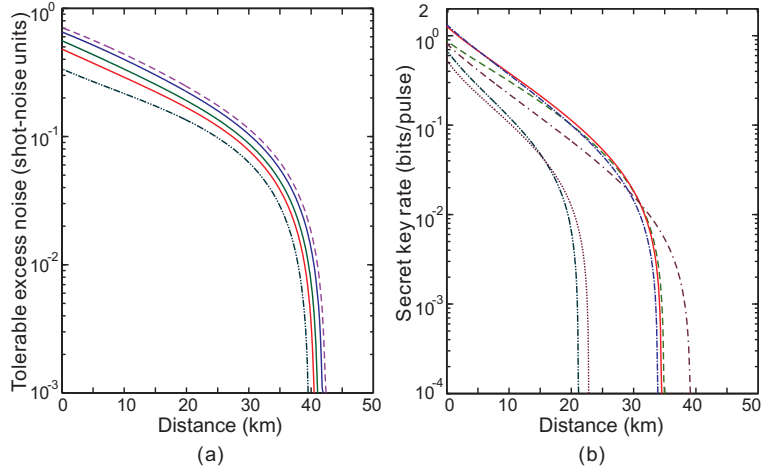,width=10cm}}
\vspace*{8pt}
\caption{(Color online) (a) Tolerable excess noise as a function of the transmission distance for $\textrm{Het}^\textit{2}$ (dashed line), $\textrm{Het}$ (dot-dot-dashed line) and $\textrm{Het}^\textit{2}_\textrm{M}$ (solid line) protocols where $T_A=0.3$ (red), 0.5 (green), 0.8 (blue) when choosing $\beta=0.99$, $V=10^5$ and $V_A=V/(1-T_A)$. (b) Secret key rate as a function of the transmission distance for $\textrm{Hom}$-$\textrm{Het}_\textrm{M}$ (dash-dash-dotted line), $\textrm{Het}$-$\textrm{Hom}_\textrm{M}$ (dashed line), $\textrm{Het}^\textit{2}_\textrm{M}$ (solid line), $\textrm{Hom}^\textit{2}_\textrm{M}$ (dash-dotted line), $\textrm{Hom}$ (dotted line) and $\textrm{Het}$ (dot-dot-dashed line) protocols when choosing $\varepsilon=0.2$, $\beta=0.99$, $T_A=0.8$, and $V_A=V=100$.}
\end{figure}

\begin{figure}[pb]
\centerline{\psfig{file=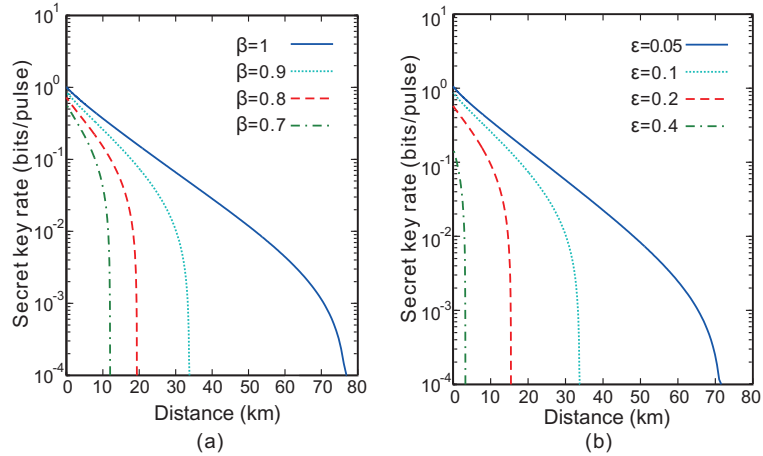,width=10cm}}
\vspace*{8pt}
\caption{(Color online) (a) Secret key rate of the new sub-protocol $\textrm{Het}^\textit{2}_\textrm{M}$  as a function of the transmission distance for $\varepsilon=0.1$ and $\beta=1, 0.9, 0.8, 0.7$. (b) Secret key rate of the new sub-protocol $\textrm{Het}^\textit{2}_\textrm{M}$  as a function of the transmission distance for $\beta=0.9$ and $\varepsilon=0.4, 0.2, 0.1, 0.05$. The curves are plotted for $T_A=0.8$ and $V=V_A=20$.}
\end{figure}

\section{Numerical Simulation and Discussion on Collective Entangling Cloner Attacks}

For simplicity in numerical simulation, we only consider for $T_1=T_2$ and $\chi_1=\chi_2$ (or $\varepsilon_1=\varepsilon_2=\varepsilon$). The tolerable excess noise $\varepsilon$ can be obtained when the secret key rate $K_R$ is zero. When $\varepsilon$, $\beta$, $T_A$, $V$ and $V_A$ are known, the elements of $\Gamma_{B_{2X}B_{2P}B_{1X}B_{1P}A_2A_{1X}A_{1P}}$ are obtained from Eq. (\ref{C1-C5}). Assuming that the typical fiber channel loss is 0.2 dB/km, with using Eq. (\ref{K_R(a_{mn})}), we numerically simulate $\varepsilon$ and $K_R$ as the functions of the transmission distance by MATLAB. For comparison, the original $\textrm{Het}^\textit{2}$ protocol,\cite{6Pirandola} the heterodyne protocol ($\textrm{Het}$) and the homodyne protocol ($\textrm{Hom}$) of one-way CV-QKD protocol\cite{het,2Grosshans02}  are also numerically simulated in Figs. 3(a) and (b), respectively.

Fig. 3(a) shows the tolerable excess noise as a function of the transmission distance for $\textrm{Het}^\textit{2}_\textrm{M}$ protocol in the case that $T_A$ changes and $V_A=V/(1-T_A)$. When choosing $\beta=0.99$, $V=10^5$ and $T_A=0.3, 0.5, 0.8$, the numerical simulation result indicates that the tolerable excess noise of $\textrm{Het}^\textit{2}_\textrm{M}$ goes up with the increase of $T_A$. $V$ and $\varepsilon$ are in shot-noise units. When $T_A$ approaches 1, the $\textrm{Het}^\textit{2}_M$ protocol asymptotically approaches the original two-way protocol $\textrm{Het}^\textit{2}$ whose tolerable excess noise surpasses that of the corresponding one-way CV-QKD protocol.\cite{6Pirandola} The other new sub-protocols also have similar numerical simulation results. Therefore, the new protocols maintain the same advantage as the original ones.

Fig. 3(b) shows the secret key rate of all the new sub-protocols as a function of the transmission distance for the noisy channel. Considering the practical scheme,\cite{detect noise J. Lodewyck,T. Symul experiment} we choose $\varepsilon=0.2$, $\beta=0.99$, $T_A=0.8$ and $V=V_A=100$. The simulation result indicates that all new protocols have higher secret key rate than the one-way CV-QKD protocols.  Note that the achievable transmission distance of $\textrm{Hom}$-$\textrm{Het}_\textrm{M}$ protocol is the longest in all the new sub-protocols. The reason is that, in $\textrm{Hom}$-$\textrm{Het}_\textrm{M}$, Bob measures  the mode $B_2$ with heterodyne detection to get the position and momentum quadratures, but only uses one of them for reconciliation. This is equivalent to Bob implementing the homodyne detection with added noise. The properly added noise is useful to enhance secret key rate.\cite{8Cerf added noiseenhance key,Pirandola added noiseenhance key,Renner added noiseenhance key,Renes added noiseenhance key}

Both Fig. 4(a) and (b) show the secret key rate of the new sub-protocol $\textrm{Het}^\textit{2}_\textrm{M}$  as a function of the transmission distance where $T_A=0.8$ and $V=V_A=20$. Fig. 4(a) is plotted for $\varepsilon=0.1$ and $\beta=1, 0.9, 0.8, 0.7$. The simulation result indicates that the secret key rate of $\textrm{Het}^\textit{2}_\textrm{M}$ protocol increases with the increase of $\beta$. Fig. 4(b) is plotted for $\beta=0.9$ and $\varepsilon=0.4, 0.2, 0.1, 0.05$. The simulation result indicates that the secret key rate of $\textrm{Het}^\textit{2}_\textrm{M}$ protocol increases with the decrease of $\varepsilon$.

\section{Conclusion}
In conclusion, we propose a family of new two-way CV-QKD protocols by replacing the displacement operation of the original two-way CV-QKD protocols with the passive operation on Alice's side. By using the optimality of Gaussian attack and the purification of the system, Eve's accessible information is bounded by the measurement values of Alice and Bob. The security of the new two-way CV-QKD protocols against collective entangling cloner attacks is proved without randomly switching between one-way and two-way schemes for the quantum-channel tomography. Thus the PM scheme of our new protocol can be applied more practically. The simulation result indicates that the tolerable excess noise in the new protocols approaches the original ones when $T_A$ is close to 1. Even if $T_A$ and $V_A$ have real experimental values, the new two-way CV-QKD protocols still outperform the one-way protocols in secret key rate and transmission distance.
Especially, the new sub-protocol $\textrm{Hom}$-$\textrm{Het}_\textrm{M}$ allows the distribution of secret keys over much longer distance than the one-way protocols. However, some open questions about the security of the new two-way CV-QKD protocols still remain. In our proof, we have not analyzed the effects  of the finite size,\cite{finite size A. Leverrier,finite size 2A. Leverrier,finite size R. Renner} the source noise\cite{source noise R. Filip,source noise R. Filip2010,source noise Ralph2010,source noise Y. Shen} and the detection noise\cite{KRB1 J. Lodewyck,detect noise J. Lodewyck} on the security. Especially, it is worthwhile to further investigate the method to enhance the tolerable excess noise of CV QKD by adding proper noise on the side of the sender or the receiver.\cite{8Cerf added noiseenhance key,Pirandola added noiseenhance key,Renner added noiseenhance key,Renes added noiseenhance key,source noise Ralph2010} These problems will be researched in our future work.

\section*{Acknowledgments}
This work is supported by the Key Project of National Natural
Science Foundation of China (Grant No. 60837004 and 61101081), National Hi-Tech Research and Development (863) Program.

\section*{Appendix A. The Equivalence of Fig. 1($\textrm{b}$) and Fig. 2 on Eve's Accessible Information}

\setcounter{equation}{0}

\renewcommand{\theequation}{A.\arabic{equation}}

In Fig. 1(b), Bob calculates two variables $x_B=x_{B_{2X}}-kx_{B_{1X}}$ and $p_B=p_{B_{2P}}+kp_{B_{1P}}$ after measuring $B_{1X}$, $B_{2X}$, $B_{1P}$ and $B_{2P}$. We name it as measure-and-calculate (MC) process. In Fig. 2, Bob measures the mode $B_4$ ($B_6$) to get the variable $x_{B_4}=x_B$ ($p_{B_6}=p_B$) after using two $\Gamma_k$ on the modes $B_{1X}$, $B_{2X}$, $B_{1P}$ and $B_{2P}$. We name it as transform-and-measure (TM) process. In the following, we prove that the two processes are equivalent for Eve's entropy $S(E)$ as well as conditional entropy $S(E|x_B,p_B)=\int_{-\infty}^\infty p(x_B,p_B)S(\rho_E^{x_B,p_B})dx_Bdp_B$, where $p(x_B,p_B)$ is the probability distribution of $x_B$ and $p_B$ and $\rho_E^{x_B,p_B}$ is Eve's state when Bob's variables $x_B$ and $p_B$ are known. We use $B_o$ to denote $B_{1X}B_{2X}B_{1P}B_{2P}$, $D$ to denote $B_3B_4B_5B_6$, and $A_o$ to denote $A_{1X}A_{1P}A_2$.

In MC process, after Bob measures  $B_{1X}$, $B_{2X}$, $B_{1P}$ and $B_{2P}$, the state $\rho_{A_oB_oE}$ is changed into $\rho_{A_oB^\prime_oE}$. Thus
\begin{equation}
\rho_{A_oB_o^\prime E}=\int_{-\infty}^\infty F_B \rho_{A_oB_oE}F_Bdx_1dx_2dp_1dp_2,
\end{equation}
where
\begin{equation}
 F_B=\left|x_1,x_2,p_1,p_2\right\rangle_{B_o}\left\langle x_1,x_2,p_1,p_2\right|.
\end{equation}
$F_B$ indicates the measurement process that obtains the corresponding eigenvalues $x_1$, $x_2$, $p_1$ and $p_2$ of $B_{1X}$, $B_{2X}$, $B_{1P}$ and $B_{2P}$.

In order to get $x_B=x_2-kx_1$ and $p_B=p_2+kp_1$, we do the parameter transformation by replacing $x_2$ and $p_2$ with $x_2=x_B+kx_1$ and $p_2=p_B-kp_1$, respectively. For the conditional state, we fix $x_B$ and $p_B$, and denote:
\begin{equation}
\rho_{A_oB_o^\prime E}^{ x_B,p_B}=\int_{-\infty}^\infty F_B^\prime \rho_{A_oB_oE}F_B^\prime dx_1dp_1,
\end{equation}
where
\begin{eqnarray}
F_B^\prime&=&\left|+-\right\rangle_{B_o}\left\langle +-\right|\nonumber \\
  &=&\left|x_1,x_B+kx_1,p_1,p_B-kp_1\right\rangle_{\!B_o}\!\left\langle x_1,x_B+kx_1,p_1,p_B-kp_1\right|.
\end{eqnarray}

When $x_B$ and $p_B$ are known, Eve's state is
\begin{eqnarray} \label{rho_E^b}
\rho_E^{x_B,p_B}\!&=&\!\frac{\textrm{tr}_{A_oB_o^\prime}\left(\rho_{A_oB_o^\prime E}^{ x_B,p_B}\right)}{\textrm{tr}_{A_oB_o^\prime E}\left(\rho_{A_oB_o^\prime E}^{ x_B,p_B}\right)} \nonumber \\
&=&\!\!\frac{\textrm{tr}_{A_o}\!\!\left(\int_{-\infty}^\infty \!_{B_o}\!\!\left\langle\! x_1^\prime,x_2^\prime,p_{\!1}^\prime,p_2^\prime\right|  \rho_{\!A_o\!B_o^\prime\! E}^{x_B,p_B}\left|x_1^\prime,x_2^\prime,p^\prime_{\!1},p_2^\prime\right\rangle\!\!_{B_o}d\!x_1^\prime d\!x_2^\prime d\!p^\prime_{\!1}d\!p_2^\prime \right)}{\textrm{tr}_{A_oE}\!\left(\int_{-\infty}^\infty \!_{B_o}\!\!\left\langle\! x_1^\prime,x_2^\prime,p^\prime_{\!1},p_2^\prime\right|  \rho_{\!A_o\!B_o^\prime\! E}^{x_B,p_B}\left|x_1^\prime,x_2^\prime,p^\prime_{\!1},p_2^\prime\right\rangle\!\!_{B_o}d\!x_1^\prime d\!x_2^\prime d\!p^\prime_{\!1}d\!p_2^\prime\!\right)} \nonumber \\
&=&\!\!\frac{\textrm{tr}_{A_o}\left(\int_{-\infty}^\infty  \!_{B_o}\!\!\left\langle +-\right| \rho_{A_oB_oE} \left|+-\right\rangle\!_{B_o}dx_1dp_1\right)}{\textrm{tr}_{A_oE}\left(\int_{-\infty}^\infty \!_{B_o}\!\!\left\langle +-\right|  \rho_{A_oB_oE}  \left|+-\right\rangle\!_{B_o}dx_1dp_1\right)}.
\end{eqnarray}

In TM process, the operation of the two unitary transformations $\Gamma_k$ is denoted as $S^T$ which can transform $\left| x_1,x_B,p_1,p_B\right\rangle_{B_o}$ into $\left| x_1,x_B+kx_1,p_1,p_B-kp_1\right\rangle_{B_o}$.\cite{QND} After implementing the two unitary transformations $\Gamma_k$, the original state $\rho_{A_oB_oE}$ is changed into $\rho_{A_oB_3B_4B_5B_6E}=S\rho_{A_oB_oE}S^T$. When getting $x_B$ and $p_B$ by measuring $B_4$ and $B_6$, the state is
\begin{equation}
\rho_{A_oB_3B_5E}^{x_B,p_B}=~_{B_4B_6}\!\left\langle x_B,p_B\right| S\rho_{A_oB_oE}S^T \left|x_B,p_B\right\rangle_{B_4B_6}.
\end{equation}

When $x_B$ and $p_B$ are known, Eve's state is
\begin{eqnarray}\label{rho E'' b}
\rho_E^{\prime x_B,p_B}\!\!\!\!&&=\!\frac{\textrm{tr}_{A_oB_3B_5}\left(\rho_{A_oB_3B_5E}^{x_B,p_B}\right)}{\textrm{tr}_{A_oB_3B_5E}\left(\rho_{A_oB_3B_5E}^{x_B,p_B}\right)} \nonumber \\
&&=\!\frac{\textrm{tr}_{A_o}\left(\int_{-\infty}^\infty  \!_{B_3B_5}\!\!\left\langle x_3,p_5\right|  \rho_{A_oB_3B_5E}^{x_B,p_B}  \left|x_3,p_5\right\rangle\!_{B_3B_5}dx_3dp_5\right)}{\textrm{tr}_{A_oE}\left(\int_{-\infty}^\infty \!_{B_3B_5}\!\!\left\langle x_3,p_5\right|  \rho_{A_oB_3B_5E}^{x_B,p_B} \left|x_3,p_5\right\rangle\!_{B_3B_5}dx_3dp_5\right)} \nonumber \\
&&=\!\frac{\textrm{tr}_{A_o}\left(\int_{-\infty}^\infty\! _D\!  \left\langle x_3,x_B,p_5,p\!_B\right|S\rho_{A_oB_oE}S^T \left|x_3,x_B,p_5,p\!_B\right\rangle\!\!_Ddx_3dp_5\right)}{\textrm{tr}_{A_oE}\left(\int_{-\infty}^\infty \!_D \!\left\langle x_3,x_B,p_5,p\!_B\right| S\rho_{A_oB_oE}S^T \! \left|x_3,x_B,p_5,p\!_B\right\rangle\!\!_Ddx_3dp_5\right)} \nonumber \\
&&=\!\frac{\textrm{tr}_{A_o}\left(\int_{-\infty}^\infty\!  _{B_o}\!\!\left\langle x_1,x_B,p_1,p\!_B\right| S\rho_{A_oB_oE}S^T \left|x_1,x_B,p_1,p\!_B\right\rangle\!\!_{B_o}dx_1dp_1\right)}{\textrm{tr}_{A_o\!E}\!\left(\int_{-\infty}^\infty \!_{B_o}\!\!\left\langle x_1,x_B,p_1,p\!_B\right|  S\rho_{A_oB_oE}S^T \! \left|x_1,x_B,p_1,p\!_B\right\rangle\!\!_{B_o}dx_1dp_1\right)} \nonumber \\
&&=\!\frac{\textrm{tr}_{A_o}\left(\int_{-\infty}^\infty\!  _{B_o}\!\!\left\langle +-\right| \rho_{A_oB_oE} \left|+-\right\rangle\!\!_{B_o}dx_1dp_1\right)}{\textrm{tr}_{A_oE}\left(\int_{-\infty}^\infty \!_{B_o}\!\!\left\langle +-\right|  \rho_{A_oB_oE} \left|+-\right\rangle\!\!_{B_o}dx_1dp_1\right)}.
\end{eqnarray}

Since Eq. (\ref{rho E'' b}) is the same as Eq. (\ref{rho_E^b}) and $P(x_B,p_B)$ is proportion to $\textrm{tr}_{A_oE}\left(\int_{-\infty}^\infty \!_{B_o}\!\!\left\langle +-\right|  \rho_{A_oB_oE} \left|+-\right\rangle\!\!_{B_o}dx_1dp_1\right)$,  $S(E|x_B,p_B)$ is identical in MC process and TM process. The cases in the other new sub-protocols can be proved in the same way.

In Fig. 2, because the state $B_2B_1A_2A_1E$ is also a pure state, $S(E)=S(B_2B_1A_2A_1)$. Similarly, in Fig. 1(b), $S(E)=S(B_2B_1A_2A_1)$. Because the modes $B_2B_1A_2A_1$ are same to Figs. 1(b) and 2, $S(E)$ is same. Therefore, $I_{BE}$ is same to Figs. 1(b) and 2.

\section*{Appendix B. The Calculation of Eq. (\ref{Gamma_C})}

\setcounter{equation}{0}

\renewcommand{\theequation}{B.\arabic{equation}}

In Fig. 2, the corresponding covariance matrixes of EPR pairs of Alice and Bob are
\begin{eqnarray}
\Gamma\!_{Bob} \!=\!\!\left(\!\!\begin{array}{cc}
   V\mathbb{I}           &  \sqrt{V^2-1}\sigma_z\\
   \sqrt{V^2-1}\sigma_z  &  V\mathbb{I}
  \end{array}\!\!\right),~~~~
\Gamma\!_{Alice} \!=\!\!\left(\!\!\begin{array}{cc}
   V_A\mathbb{I}           &  \sqrt{V_A^2-1}\sigma_z\\
   \sqrt{V_A^2-1}\sigma_z  &  V_A\mathbb{I}
  \end{array}\!\!\right).
\end{eqnarray}
The two modes $B_1$ and $A_1$ are uncorrelated. The mode $C_1$ is changed into the mode $A_{in}$ through the channel. Alice couples one mode of her EPR pair with the mode $A_{in}$ by the beam splitter $T_A$. The action of the beam splitter $T_A$ is equivalent to a unitary transformation. When the mode $A_{out}$ is sent back to Bob, the corresponding covariance matrix of the modes $B_2B_1A_2A_1$ is
\begin{eqnarray}
\Gamma_{\!\!B\!_2\!B\!_1\!A\!_2\!A\!_1}\!\!\! =\!\!\!\left(\!\!\!\begin{array}{cccccccc}
   V_{\!x\!_{B_2}}& 0& C_{\!x\!_{B_2}x_{\!B\!_1}}& 0& C_{\!x\!_{B_2}x_{\!A\!_2}}& 0& C_{\!x\!_{B_2}x_{\!A\!_1}}& 0\\
   0& V_{\!p\!_{B_2}}& 0& C_{\!p\!_{B_2}p_{\!B\!_1}}& 0& C_{\!p\!_{B_2}p_{\!A\!_2}}& 0& C_{\!p\!_{B_2}p_{\!A\!_1}}\\
   C_{\!x\!_{B_2}x_{\!B\!_1}}& 0& V& 0& C_{\!x_{\!B\!_1}x_{\!A\!_2}}& 0& 0& 0\\
   0& C_{\!p\!_{B_2}p_{\!B\!_1}}& 0& V& 0& C_{\!p_{\!B\!_1}p_{\!A\!_2}}& 0& 0\\
   C_{\!x\!_{B_2}x_{\!A\!_2}}& 0& C_{\!x_{\!B\!_1}x_{\!A\!_2}}& 0& V_{\!x_{\!A\!_2}}& 0& C_{\!A_1\!A_2}& 0\\
   0& C_{\!p\!_{B_2}p_{\!A\!_2}}& 0& C_{\!p_{\!B\!_1}p_{\!A\!_2}}& 0& V_{\!p_{\!A\!_2}}& 0& -C_{\!A\!_1\!A_2}\\
   C_{\!x\!_{B_2}x_{\!A\!_1}}& 0& 0& 0& C_{\!A_1\!A_2}& 0& V_{\!A}& 0\\
   0& C_{\!p\!_{B_2}p_{\!A\!_1}}& 0& 0& 0& -C_{\!A\!_1\!A_2}& 0& V_{\!A}
  \end{array}\!\!\!\!\right),
\end{eqnarray}
where the diagonal elements correspond to the variances of $x$ and $p$ quadratures of the modes $B_2$, $B_1$, $A_2$ and $A_1$ in turn, and the nondiagonal elements correspond to the covariances between modes. Note that the covariance between the modes $A_1$ and $A_2$ is $C_{A_1A_2}=\sqrt{T_{\!A}\left(V_{\!A}^2 - 1\right)}$, which is irrelevant to the channels since the mode $A_1$ is only controlled by Alice and its values are random.

In the heterodyne detection, a vacuum state is introduced by the beam splitter. The corresponding covariance matrix of the modes $B_2B_1A_2A_1$ and the three vacuum states $C_{01}$, $C_{02}$ and $C_{03}$ is
\begin{eqnarray} \label{B3}
&&\Gamma_{B_2C_{01}B_1C_{02}A_2A_1C_{03}} =\nonumber\\
&&\left(\!\begin{array}{cccccccccccccc}
   V_{\!x\!_{B_2}}& 0&0&0& C_{\!x\!_{B_2}x_{\!B\!_1}}& 0&0&0& C_{\!x\!_{B_2}x_{\!A\!_2}}& 0& C_{\!x\!_{B_2}x_{\!A\!_1}}& 0&0&0\\
   0& V_{\!p\!_{B_2}}&0&0& 0& C_{\!p\!_{B_2}p_{\!B\!_1}}&0&0& 0& C_{\!p\!_{B_2}p_{\!A\!_2}}& 0& C_{\!p\!_{B_2}p_{\!A\!_1}}&0&0\\
   0&0&1&0&0&0&0&0&0&0&0&0&0&0\\
   0&0&0&1&0&0&0&0&0&0&0&0&0&0\\
   C_{\!x\!_{B_2}x_{\!B\!_1}}& 0&0&0& V& 0&0&0& C_{\!x_{\!B\!_1}x_{\!A\!_2}}& 0& 0& 0&0&0\\
   0& C_{\!p\!_{B_2}p_{\!B\!_1}}&0&0& 0& V&0&0& 0& C_{\!p_{\!B\!_1}p_{\!A\!_2}}& 0& 0&0&0\\
   0&0&0&0&0&0&1&0&0&0&0&0&0&0\\
   0&0&0&0&0&0&0&1&0&0&0&0&0&0\\
   C_{\!x\!_{B_2}x_{\!A\!_2}}& 0&0&0& C_{\!x_{\!B\!_1}x_{\!A\!_2}}& 0&0&0& V_{\!x_{\!A\!_2}}& 0& C_{A_1A_2}& 0&0&0\\
   0& C_{\!p\!_{B_2}p_{\!A\!_2}}&0&0& 0& C_{\!p_{\!B\!_1}p_{\!A\!_2}}&0&0& 0& V_{\!p_{\!A\!_2}}& 0& -C_{A_1A_2}&0&0\\
   C_{\!x\!_{B_2}x_{\!A\!_1}}& 0&0&0& 0& 0&0&0& C_{A_1A_2}& 0& V_{\!A}& 0&0&0\\
   0& C_{\!p\!_{B_2}p_{\!A\!_1}}&0&0& 0& 0&0&0& 0& -C_{A_1A_2}& 0& V_{\!A}&0&0\\
   0&0&0&0&0&0&0&0&0&0&0&0&1&0\\
   0&0&0&0&0&0&0&0&0&0&0&0&0&1\\
  \end{array}\right). \nonumber\\
\end{eqnarray}
By the unitary transformations of the three beam splitters, the modes $B_2B_1A_2A_1$ are changed into the modes $B_{2X}B_{2P}B_{1X}B_{1P}A_2A_{1X}A_{1P}$. Its corresponding covariance matrix is $\Gamma_{B_{2X}B_{2P}B_{1X}B_{1P}A_2A_{1X}A_{1P}}
=[\Gamma_{\textrm{BS}}\oplus\Gamma_{\textrm{BS}}\oplus\mathbb{I}\oplus\Gamma_{\textrm{BS}}]\Gamma_{B_2C_{01}B_1C_{02}A_2A_1C_{03}}[\Gamma_{\textrm{BS}}\oplus\Gamma_{\textrm{BS}}\oplus\mathbb{I}\oplus\Gamma_{\textrm{BS}}]^T$, where
\begin{equation}
  \Gamma_{\textrm{BS}}=\left(
\begin{array}{cccc}
\sqrt{\frac{1}{2}}&0&\sqrt{\frac{1}{2}}&0\\
0&\sqrt{\frac{1}{2}}&0&\sqrt{\frac{1}{2}}\\
-\sqrt{\frac{1}{2}}&0&\sqrt{\frac{1}{2}}&0\\
0&-\sqrt{\frac{1}{2}}&0&\sqrt{\frac{1}{2}}
\end{array}
\right).
\end{equation}
Therefore, Eq. (\ref{Gamma_C}) is obtained, in which
\begin{eqnarray}
  && \gamma_{B_{2X}}\! \!=\!\gamma_{B_{2P}}\!=\!\left(\!\begin{array}{cc}
\frac{1\! +\! V_{\!x\!_{B_2}}}{2}& 0\\
 0& \frac{1\! +\! V_{\!p\!_{B_2}}}{2}
  \end{array}\!\right),~~~~~
  \gamma_{A_2} =\left(\begin{array}{cc}
  V_{\!x_{\!A\!_2}}& 0\\
 0& V_{\!p_{\!A\!_2}}
  \end{array}\right),\nonumber\\
  &&C_1 =\left(\begin{array}{cc}
  \frac{C_{\!x\!_{B_2}x_{\!B\!_1}}}{2}& 0\\
 0& \frac{C_{\!p\!_{B_2}p_{\!B\!_1}}}{2}
  \end{array}\right),~~~~~~~~~~~~~
  C_2 =\left(\begin{array}{cc}
  \frac{C_{\!x\!_{B_2}x_{\!A\!_2}}}{\sqrt{2}}& 0\\
 0& \frac{C_{\!p\!_{B_2}p_{\!A\!_2}}}{\sqrt{2}}
  \end{array}\right),\nonumber\\
  &&C_3 =\left(\begin{array}{cc}
  \frac{C_{\!x\!_{B_2}x_{\!A\!_1}}}{2}& 0\\
 0& \frac{C_{\!p\!_{B_2}p_{\!A\!_1}}}{2}
  \end{array}\right),~~~~~~~~~~~~~
  C_4 =\left(\begin{array}{cc}
  \frac{C_{\!x_{\!B\!_1}x_{\!A\!_2}}}{\sqrt{2}}& 0\\
 0& \frac{C_{\!p_{\!B\!_1}p_{\!A\!_2}}}{\sqrt{2}}
  \end{array}\right),
  \nonumber\\
  &&C_5 \!=\!\left(\!\!\begin{array}{cc}
  \!\sqrt{\frac{\!T_{\!A}\left(\!V_{\!A}^2\! - 1\!\right)}{2}}& 0\\
 0& \!-\!\sqrt{\frac{\!T_{\!A}\left(\!V_{\!A}^2\! - 1\!\right)}{2}}
  \end{array}\!\right).
\end{eqnarray}

Every element of Eq. (\ref{Gamma_C}) can be obtained by the measurement values in experiment. For example, in the heterodyne detection on the mode $B_2$, the $x$ quadrature value $x_{B_{2X}}$ of the mode $B_{2X}$ and the $p$ quadrature value $p_{B_{2P}}$ of the mode $B_{2P}$ are obtained. There are the following relations
\begin{eqnarray}
  &&x_{B_{2X}} =\sqrt{\frac{1}{2}}(x_{B_2}+x_0)  ,~~~~~~
  x_{B_{2P}} =\sqrt{\frac{1}{2}}(x_0-x_{B_2}),\nonumber\\
  &&p_{B_{2X}} =\sqrt{\frac{1}{2}}(p_{B_2}+p_0)  ,~~~~~~
  p_{B_{2P}} =\sqrt{\frac{1}{2}}(p_0-p_{B_2}),
\end{eqnarray}
where $x_{B_2}$, $p_{B_2}$, $x_0$ and $p_0$ are the $x$ and $p$ quadratures of the mode $B_2$ and the vacuum state, respectively, $p_{B_{2X}}$ is the $p$ quadrature of the mode $B_{2X}$ and $x_{B_{2P}}$ is the $x$ quadrature of the mode $B_{2P}$. Then, we get
\begin{eqnarray}
  &&p_{B_{2X}} =-p_{B_{2P}}+\sqrt{2}p_0,\nonumber\\
  &&x_{B_{2P}} =-x_{B_{2X}}+\sqrt{2}x_0.
\end{eqnarray}
Therefore, the variances of $p$ and $x$ quadratures of the modes $B_{2X}$ and $B_{2P}$ can be calculated according to the measurement values $x_{B_{2X}}$ and $p_{B_{2P}}$
\begin{eqnarray}
  &&\left\langle p_{B_{2X}}^2\right\rangle =\left\langle p_{B_{2P}}^2\right\rangle-2\sqrt{2}\left\langle p_{B_{2P}}p_0\right\rangle+2\left\langle p_0^2 \right\rangle=\left\langle p_{B_{2P}}^2\right\rangle,\nonumber\\
  &&\left\langle x_{B_{2P}}^2\right\rangle=\left\langle x_{B_{2X}}^2\right\rangle- 2\sqrt{2}\left\langle x_{B_{2X}}x_0\right\rangle+2\left\langle x_0^2\right\rangle =\left\langle x_{B_{2X}}^2\right\rangle.
\end{eqnarray}

Similarly, the covariances between modes can be calculated. For example,
\begin{eqnarray}
  C_2 &=&diag\left(\left\langle x_{B_{2X}} x_{A_2} \right\rangle,\left\langle p_{B_{2X}}p_{A_2}\right\rangle\right)\nonumber\\
  &=&diag\left(\left\langle x_{B_{2X}} x_{A_2} \right\rangle,\left\langle (-p_{B_{2P}}+\sqrt{2}p_0)p_{A_2}\right\rangle\right)\nonumber\\
  &=&diag\left(\left\langle x_{B_{2X}} x_{A_2} \right\rangle,\left\langle -p_{B_{2P}}p_{A_2}\right\rangle\right),
\end{eqnarray}
where $x_{A_2}$ and $p_{A_2}$ are the measurement values of $x$ and $p$ quadratures of the mode $A_2$ which are obtained by randomly measuring the $x$ and $p$ quadratures of the mode $A_2$.

\section*{Appendix C. The Calculation of Eigenvalues}

\setcounter{equation}{0}

\renewcommand{\theequation}{C.\arabic{equation}}

The corresponding covariance matrix $\Gamma$ of a $n$-mode state has $n$ eigenvalues $\lambda_i^{\prime\prime}$ for $i=1,...,n$ where $\lambda_i^{\prime\prime}$ is the function of the element $\alpha^{\prime\prime}_{mn}$ of $\Gamma$.
The symplectic invariants of the n-mode state $\{\vartriangle_{n,j}\}$ for $j=1,..., n$ are defined as\cite{eigenvalue}
\begin{equation}
  \vartriangle_{n,j}=M_{2j}(\Omega\Gamma),
\end{equation}
where $\Omega=\oplus_1^ni\sigma_y$ ($\sigma_y$ standing for the $y$ Pauli matrix) and $M_{2j}(\Omega\Gamma)$ is the principal minor of order $2j$ of the $2n\times2n$ matrix $\Omega\Gamma$ which is the sum of the determinants of all the $2j\times2j$  submatrices of $\Omega\Gamma$ obtained by deleting $2n\!-\!2j$ rows and the
corresponding $2n\!-\!2j$ columns.\cite{eigenvalue} There are $n$ independent symplectic invariants $\{\!\vartriangle\!\!\!_{n,j}\!\}$ which are the function of the element $\alpha^{\prime\prime}_{mn}$ of $\Gamma$. In addition, there is a relation\cite{eigenvalue}
 \begin{equation}
  \vartriangle_{n,j}=\sum_{s_j^n}\prod_{i\in s_j^n}{\lambda_i^{\prime\prime}}^2,
\end{equation}
where $s_j^n$ are the subsets of all the possible combinations of $j$ integers within $n$ where $j$ is smaller than or equal to $n$. Therefore, the symplectic eigenvalues $\lambda_i^{\prime\prime}$ for $i=1,...,n$ are the solutions of the $n$ order polynomial
\begin{equation}
  z^n-\vartriangle_{n,1}z^{n-1}+\vartriangle_{n,2}z^{n-2}-\vartriangle_{n,3}z^{n-3}+...\vartriangle_{n,n}=0.
\end{equation}
The solutions are denoted as $z=(\lambda_i^{\prime\prime})^2=f^2_{\lambda_i^{\prime\prime}}(\alpha^{\prime\prime}_{mn})$ for $i=1,...,n$ which are the function of the element $\alpha^{\prime\prime}_{mn}$ of the covariance matrix $\Gamma$. For $n=4$, there are
\begin{eqnarray} \label{lamda1234}
  f^2\!\!\!\!_{\lambda_{1,2}^{\prime\prime}}\!(\!\alpha^{\prime\prime}_{mn}\!)\!\!=\!\!\frac{\vartriangle_{4\!,1}}{4}\!-\!\frac{1}{2}\sqrt{\!\frac{\vartriangle_{4\!,1}^2}{4}\!-\!\frac{2\!\!\vartriangle_{4\!,2}}{3}\!+\!\Theta}\pm\frac{1}{2}\!\sqrt{\!\!\frac{\vartriangle_{4\!,1}^2}{2}\!-\!\frac{4\!\!\vartriangle_{4\!,2}}{3}\!-\!\Theta\!-\!\frac{\vartriangle_{4\!,1}^3 \!- 4 \!\!\vartriangle_{4\!,1} \vartriangle_{4\!,2} \!+\! 8 \!\!\vartriangle_{4\!,3}}{4\sqrt{\!\frac{\vartriangle_{4\!,1}^2}{4}\!-\!\frac{2\vartriangle_{4\!,2}}{3}\!+\!\Theta}}} , \nonumber\\
   f^2\!\!\!\!_{\lambda_{3,4}^{\prime\prime}}\!(\!\alpha^{\prime\prime}_{mn}\!)\!\!=\!\!\frac{\vartriangle_{4\!,1}}{4}\!+\!\frac{1}{2}\sqrt{\!\frac{\vartriangle_{4\!,1}^2}{4}\!-\!\frac{2\!\!\vartriangle_{4\!,2}}{3}\!+\!\Theta}\pm\frac{1}{2}\!\sqrt{\!\!\frac{\vartriangle_{4\!,1}^2}{2}\!-\!\frac{4\!\!\vartriangle_{4\!,2}}{3}\!-\!\Theta\!+\!\frac{\vartriangle_{4\!,1}^3 \!- 4 \!\!\vartriangle_{4\!,1} \vartriangle_{4\!,2} \!+\! 8\!\!\vartriangle_{4\!,3}}{4\sqrt{\!\frac{\vartriangle_{4\!,1}^2}{4}\!-\!\frac{2\vartriangle_{4\!,2}}{3}\!+\!\Theta}}},\nonumber\\
\end{eqnarray}
where
\begin{eqnarray}
  &&\Theta=\frac{2^{\frac{1}{3}}H}{3J}+\frac{J}{3\cdot 2^{\frac{1}{3}}},  \nonumber\\
  &&H=\vartriangle_{4,2}^2 - 3 \vartriangle_{4,1} \vartriangle_{4,3} + 12 \vartriangle_{4,4},   \nonumber\\
  &&J = \left(L+\sqrt{L^2-4 H^3}\right)^{\frac{1}{3}},  \nonumber\\
  &&L=2 \vartriangle_{4,2}^3 - 9 \vartriangle_{4,1} \vartriangle_{4,2} \vartriangle_{4,3} + 27 \vartriangle_{4,3}^2 + 27 \vartriangle_{4,1}^2 \vartriangle_{4,4} - 72 \vartriangle_{4,2} \vartriangle_{4,4}.
\end{eqnarray}

By a unitary transformation, $\Gamma_{AB}$ can be changed into Eq. (\ref{B3}), i.e., $diag(\Gamma_{\!B_2B_1A_2A_1},\mathbb{I}_3)$. Therefore, the eigenvalues of $\Gamma_{AB}$ are $\lambda_i=f_{\lambda_{1,2,3,4}}(\alpha_{mn}),1,1,1$, where $f_{\lambda_{1,2,3,4}}(\alpha_{mn})$ are the eigenvalues of $\Gamma_{\!B_2B_1A_2A_1}$ calculated according to Eq. (\ref{lamda1234}). By the unitary of the beam splitter, there is $[\mathbb{I}_3\oplus\Gamma_{\textrm{BS}}]^T\Gamma_{B_3B_5A_2A_{1X}A_{1P}}^{x_B,p_B}[\mathbb{I}_3\oplus\Gamma_{\textrm{BS}}]=diag(\Gamma_{B_3B_5A_2A_1}^{x_B,p_B},\mathbb{I})$, where $\Gamma_{B_3B_5A_2A_1}^{x_B,p_B}$ is the corresponding covariance matrix of a four-mode state.
Thus, the eigenvalues of $\Gamma_{B_3B_5A_2A_{1X}A_{1P}}^{x_B,p_B}$ are $\lambda^\prime_i=f_{\lambda^\prime_{1,2,3,4}}(\alpha^\prime_{mn}),1$, where $f_{\lambda^\prime_{1,2,3,4}}(\alpha^\prime_{mn})$ are the eigenvalues of $\Gamma_{B_3B_5A_2A_1}^{x_B,p_B}$ calculated according to Eq. (\ref{lamda1234}).

\section*{Appendix D. The Secret Key Rate of the $\textrm{Hom}_\textrm{M}^\textit{2}$, $\textrm{Hom}$-$\textrm{Het}_\textrm{M}$ and $\textrm{Het}$-$\textrm{Hom}_\textrm{M}$ Protocols}

\setcounter{equation}{0}

\renewcommand{\theequation}{D.\arabic{equation}}

In Fig. 2, because $S(E)=S(B_2B_1A_2A_1)$ and the modes $B_2B_1A_2A_1$ are same to all the new two-way sub-protocols, $S(E)$ is same. Therefore, we only need to consider the conditional entropy on Bob to calculate $I_{BE}$.

In $\textrm{Hom}^\textit{2}_\textrm{M}$ protocol, Bob gets the variables $x_{B_1}$ and $x_{B_2}$ by homodyne detection on the modes $B_1$ and $B_2$ and uses $x_B^\prime=x_{B_2}-kx_{B_1}$ for postprocessing. This procedure is equivalent to the one where Bob uses $\Gamma_k$ to change the modes $B_1$ and $B_2$ into $B_3^\prime$ and $B_4^\prime$. The corresponding covariance matrix of the system $B_4^\prime B_3^\prime A_o$ is
\begin{equation}
\Gamma_{B_4^\prime B_3^\prime A_o}=\left[\Gamma_k\oplus\mathbb{I}_3\right]\Gamma_{B_2B_1A_o}\left[\Gamma_k\oplus\mathbb{I}_3\right]^T,
\end{equation}
where $\Gamma_{B_2B_1A_o}$ is obtained by applying the unitary transformation $[\Gamma_{\textrm{BS}}\oplus\Gamma_{\textrm{BS}}\oplus\mathbb{I}_3]$ to Eq. (\ref{Gamma_C}).

When Bob gets the $x_B^\prime$ by measuring $B_4^\prime$, the state $B_3^\prime A_oE$ is a pure state, which means $S(E|x_B^\prime)=S(B_3^\prime A_o|x_B^\prime)$. Similar to  Eq. (\ref{SE}), we get
\begin{equation}
S(E|x_B^\prime)=\sum_{i=1}^4 G(\lambda_j^\prime),
\end{equation}
where $\lambda_j^\prime$ is the symplectic eigenvalue of the corresponding covariance matrix $\Gamma_{B_3^\prime A_o}^{x_B^\prime}$ of the state $B_3^\prime A_o$ conditioned  on $x_B^\prime$. $\Gamma_{B_3^\prime A_o}^{x_B^\prime}$ is calculated from $\Gamma_{B_4^\prime B_3^\prime A_o}$.\cite{doctor Raul,J. Eisert conditional matrix}

In $\textrm{Hom}$-$\textrm{Het}_\textrm{M}$ protocol, Bob gets the variable $x_{B_1}$ by homodyne detection on $B_1$ and gets the variables $x_{B_{2X}}$ and $p_{B_{2P}}$ by heterodyne detection on $B_2$. Bob only uses $x_B^{\prime\prime}=x_{B_{2X}}-kx_{B_1}$ for postprocessing. This procedure is equivalent to the one where Bob uses $\Gamma_k$ to change the modes $B_{2X}$ and $B_1$ into $B_3^{\prime\prime}$ and $B_4^{\prime\prime}$. The corresponding matrix of the state $B_4^{\prime\prime}B_3^{\prime\prime}B_{2p}A_o$ is
\begin{equation}
\Gamma_{B_4^{\prime\prime}B_3^{\prime\prime}B_{2p}A_o}=[\Gamma_k\oplus\mathbb{I}_4]\Gamma_{B_{2X}B_1B_{2P}A_o}[\Gamma_k\oplus\mathbb{I}_4]^T,
\end{equation}
where $\mathbb{I}_4=\mathbb{I}_3\oplus\mathbb{I}$ and $\Gamma_{B_{2X}B_1B_{2P}A_o}$ is obtained by applying the unitary transformation $[\mathbb{I}\oplus\mathbb{I}\oplus\Gamma_{\textrm{BS}}\oplus\mathbb{I}_3]$ to Eq. (\ref{Gamma_C}).

When Bob gets the variable $x_B^{\prime\prime}$ by measuring $B_4^{\prime\prime}$, the state $B_3^{\prime\prime}B_{2p}A_oE$ is a pure state, which means $S(E|x_B^{\prime\prime})=S(B_3^{\prime\prime}B_{2p}A_o|x_B^{\prime\prime})$. Similar to  Eq. (\ref{SE}), we can get
\begin{equation}
S(E|x_B^{\prime\prime})=\sum_{i=1}^5 G(\lambda_j^{\prime\prime}),
\end{equation}
where $\lambda_j^{\prime\prime}$ is the symplectic eigenvalue of the corresponding covariance matrix $\Gamma_{B_3^{\prime\prime}B_{2p}A_o}^{x_B^{\prime\prime}}$ of the state $B_3^{\prime\prime}B_{2p}A_o$ conditioned  on $x_B^{\prime\prime}$. $\Gamma_{B_3^{\prime\prime}B_{2p}A_o}^{x_B^{\prime\prime}}$ is calculated from $\Gamma_{B_4^{\prime\prime}B_3^{\prime\prime}B_{2p}A_o}$.\cite{doctor Raul,J. Eisert conditional matrix}

In $\textrm{Het}$-$\textrm{Hom}_\textrm{M}$  protocol, Bob gets the variables $x_{B_{1X}}$ and $p_{B_{1P}}$ by heterodyne detection on $B_1$ and gets the variable $x_{B_2}$ by homodyne detection on $B_2$. Bob only uses $x_B^{\prime\prime\prime}=x_{B_2}-kx_{B_{1X}}$ for postprocessing. This procedure is equivalent to the one where Bob uses $\Gamma_k$ to change the modes $B_{1X}$ and $B_2$ into $B_3^{\prime\prime\prime}$ and $B_4^{\prime\prime\prime}$. The corresponding matrix of the state $B_4^{\prime\prime\prime}B_3^{\prime\prime\prime}B_{1p}A_o$ is
\begin{equation}
\Gamma_{B_4^{\prime\prime\prime}B_3^{\prime\prime\prime}B_{1p}A_o}=[\Gamma_k\oplus\mathbb{I}_4]\Gamma_{B_2B_{1X}B_{1P}A_o}[\Gamma_k\oplus\mathbb{I}_4]^T,
\end{equation}
where $\Gamma_{B_2B_{1X}B_{1P}A_o}$ is obtained by applying the unitary transformation $[\Gamma_{\textrm{BS}}\oplus\mathbb{I}_4\oplus\mathbb{I}]$ to Eq. (\ref{Gamma_C}).

When Bob gets the variable $x_B^{\prime\prime\prime}$ by measuring $B_4^{\prime\prime\prime}$, the state $B_3^{\prime\prime\prime}B_{1P}A_oE$ is a pure state, which means $S(E|x_B^{\prime\prime\prime})=S(B_3^{\prime\prime\prime}B_{1P}A_o|x_B^{\prime\prime\prime})$. Similar to  Eq. (\ref{SE}), we can get
\begin{equation}
S(E|x_B^{\prime\prime\prime})=\sum_{i=1}^5 G(\lambda_j^{\prime\prime\prime}),
\end{equation}
where $\lambda_j^{\prime\prime\prime}$ is the symplectic eigenvalue of the corresponding covariance matrix $\Gamma_{B_3^{\prime\prime\prime}B_{1P}A_o}^{x_B^{\prime\prime\prime}}$ of the state $B_3^{\prime\prime\prime}B_{1P}A_o$ conditioned  on $x_B^{\prime\prime\prime}$. $\Gamma_{B_3^{\prime\prime\prime}B_{1P}A_o}^{x_B^{\prime\prime\prime}}$ is calculated from $\Gamma_{B_4^{\prime\prime\prime}B_3^{\prime\prime\prime}B_{1P}A_o}$.\cite{doctor Raul,J. Eisert conditional matrix}

In addition, we can obtain that, in $\textrm{Hom}^\textit{2}_\textrm{M}$ protocol,
\begin{equation}
I_{BA}=\frac{1}{2} \log_2\frac{V_A - T_AV_A  + T_AT_1 F+ \chi_2}{1  -T_A + T_AT_1 F+ \chi_2},
\end{equation}
in $\textrm{Hom}$-$\textrm{Het}_\textrm{M}$ protocol,
\begin{equation}
I_{BA}=\frac{1}{2} \log_2 \frac{1 + T_1T_2T_A F + T_2(V_A - T_AV_A + \chi_2)}{1 + T_1T_2T_A F + T_2(1 - T_A + \chi_2)},
\end{equation}
and in $\textrm{Het}$-$\textrm{Hom}_\textrm{M}$  protocol,
\begin{equation}
I_{BA}=\frac{1}{2} \log_2\frac{V_A - T_AV_A  + T_AT_1(1+ F)+ \chi_2}{1 -T_A + T_AT_1(1+ F)+ \chi_2}.
\end{equation}

According to Eq. (\ref{KR}), the secret key rate of above sub-protocols can be obtained.

\vspace*{-6pt}   %ONLY NECESSARY

%%PLS. NOTE TEMPLATE HERE SHOWN FOR DIFFERENT STYLES FOR
%%TEXT IN JOURNALS, BOOKS, REVIEW VOLUME AND PROCS. ETC.
%%AT ACTUAL WORK PLS. TYPE WITHOUT THE \myhead COMMAND.
%%YOU CAN FOLLOW MY EXAMPLE FOR THE COMMAND OF
%%\thebibliography AFTER THE \end{document} COMMAND.

\vspace*{-5pt}   %ONLY NECESSARY

\end{document}